# RadonPy: Automated Physical Property Calculation using All-atom Classical Molecular Dynamics Simulations for Polymer Informatics

Yoshihiro Hayashi,[1,2]* Junichiro Shiomi,[1,2] Junko Morikawa,[1,3] Ryo Yoshida[1,4,5]*

[1] Data Science Center for Creative Design and Manufacturing, The Institute of Statistical Mathematics (ISM), Research Organization of Information and Systems, 10-3 Midori-cho, Tachikawa, Tokyo 190-8562, Japan

[2] Department of Mechanical Engineering, The University of Tokyo, 7-3-1 Hongo, Bunkyo, Tokyo 113-8656, Japan

[3] Department of Materials Science and Engineering, School of Materials and Chemical Technology, Tokyo Institute of Technology, 2-12-1-E4-6 Ookayama, Meguro-ku, Tokyo 152-8552, Japan

[4] Research and Services Division of Materials Data and Integrated System (MaDIS), National Institute for Materials Science (NIMS), 1-2-1 Sengen, Tsukuba, Ibaraki 305-0047, Japan

[5] Department of Statistical Science, School of Multidisciplinary Science, The Graduate University of Advanced Studies (SOKENDAI), 10-3 Midori-cho, Tachikawa, Tokyo 190-8562, Japan

* To whom correspondence should be addressed. E-mail: (Y.H.) yhayashi@ism.ac.jp, (R.Y.) yoshidar@ism.ac.jp

ABSTRACT

The rapid growth of data-driven materials research has made it necessary to develop systematically designed, open databases of material properties. However, there are few open databases for polymeric materials compared to other material systems such as light-weight organic molecules and inorganic crystals. To this end, we developed RadonPy, the world's only open-source library for fully automated all-atom classical molecular dynamics (MD) simulations.

For a given polymer repeating unit, the entire process of molecular modeling, equilibrium and nonequilibrium MD calculations, and property calculations can be conducted fully automatically. In this study, 15 different properties, including the thermal conductivity, density, specific heat capacity, thermal expansion coefficients, bulk modulus, and refractive index, were calculated for more than 1,000 unique amorphous polymers. The calculated properties were compared and validated systematically with experimental values from PoLyInfo, the world's largest database of polymer properties. During the high-throughput data production, eight amorphous polymers with extremely high thermal conductivities, exceeding 0.4 W/m·K, were identified, including six polymers with unreported thermal conductivities. These polymers were found to have a high density of hydrogen bonding units or rigid, linear backbones. In addition, a decomposition analysis of the heat conduction, which is implemented in RadonPy, revealed the underlying mechanisms that yield a high thermal conductivity of the amorphous polymers: heat transfer via hydrogen bonds and dipole–dipole interactions between the polymer chains with their hydrogen bonding units or via the covalent bonds of the polymer backbone with high rigidity and linearity. The creation of massive amounts of computational property data using RadonPy will facilitate the development of polymer informatics, similar to how the emergence of the first-principles computational database for inorganic crystals had significantly advanced materials informatics.

**1. INTRODUCTION**

Materials informatics (MI) is a growing interdisciplinary field of materials science, attracting significant attention in recent years. MI utilizes machine learning to model, predict, and optimize the properties of new materials.[1,2] Naturally, the most essential resource in MI is data. Hence, significant efforts have been made to develop open databases for inorganic materials and light-weight organic molecules, such as the Materials Project[3] (~140,000 inorganic compounds), the Automatic-Flow[4] (AFLOW: ~3,000,000 inorganic compounds), the Open Quantum Materials Database[5] (OQMD:~1,000,000 inorganic compounds), and QM9[6] (~134,000 organic molecules). In particular, the huge databases of computational properties built using high-throughput first-principles calculations have brought remarkable progress in MI and their widespread use in science and technology. However, for polymeric materials, despite their industrial usefulness and unique characteristics, such as lightness, high tenacity, elasticity, and ease of processing, the development of open databases has considerably lagged behind other material systems.[7] This is due to the following reasons: (1) high costs of data production, (2) the difficulty in creating common data due to the diversity of polymeric materials in terms of structures and processing conditions, and (3) cultural barriers to avoiding information leakage to competitors.[2] In addition, the computational difficulty in performing high-throughput calculations and their high computational costs have hindered the development of computational property databases for polymeric materials.

PoLyInfo[8] is the current largest database of polymer properties, built from manually collected literature data. Currently, it contains approximately 100 properties of more than 18,000 polymers. However, the overall data in PoLyInfo are rather sparse as there are few cases where more than one property is simultaneously recorded for one polymer. Polymer Genome[9–12] is a database of crystalline polymers, constructed using first-principles calculations. It contains seven different electronic and optical properties, including the crystal bandgap (562 polymers), polymer chain bandgap (3,881 polymers), static dielectric constant of polymer crystals (383 polymers), and refractive index of polymer crystals (383 polymers). A common feature of these databases is that they do not provide application programming interfaces (APIs) and therefore do not allow automatic batch downloading of the data. Therefore, the creation of data resources conducive to data-driven research is vital for advancing polymer informatics.

Large-scale data of computational properties have proven to be an essential resource for machine learning applications in MI. For example, such big data have been used as source data for transfer learning when dealing with limited data in materials research. Transfer learning represents a statistical methodology for reusing knowledge, data, or models acquired in one domain (source domain) to another (target domain).[13,14] Suppose that directly establishing a machine learning model from scratch is difficult due to the lack of sufficient amount of experimental data, in such cases, a model is trained on a large amount of computational property data, and the pretrained model is fine-tuned using a small amount of experimental data, to build a highly accurate prediction model in the target domain. Successful examples of cross-domain transfer between computational and experimental data have been reported for various material systems,[15–20] including our previous work on the prediction and synthesis of thermally conductive amorphous polymers using neural networks transferred from computational properties in which only 28 samples were available in the target domain.[21]

For polymer properties, even computational data are rather scarce. Polymer Genome[9–12] is the only existing database, which is constructed using first-principles electronic structure calculations of polymers in crystalline states. However, currently, the number of samples is small, and the calculation is limited to seven electrical and optical properties. Afzal et al. created a dataset of 315 polymers using high-throughput molecular dynamics (MD) simulations; however, the target properties were limited to the glass transition temperature ($T_g$) and thermal expansion coefficient.[22] To build a computational polymer property database, a workflow of high-throughput MD simulations should be established, which is considered technically challenging. The entire workflow of an MD simulation comprises several sub-modules, such as the specification of an empirical potential, the initialization of polymer chains, equilibrium and nonequilibrium MD simulations, and the calculation of the properties from simulated molecular trajectories, which complicate the job control and error handling when fully automating the workflow. In addition, various types of conditional parameters, such as the degree of polymerization, number of polymer chains, and annealing schedules, need to be determined appropriately. Furthermore, a unified platform is required to create various polymeric states such as amorphous structures, oriented structures, and polymer blends. It also requires vast computational resources. For example, an equilibrium MD simulation of a conventional amorphous polymer requires an average run time of more than 30–50 h based on our experiments conducted on a PC with two CPU (Intel Xeon Gold 6148; 2.4 GHz) having 40 cores.

Herein, we present RadonPy (https://github.com/RadonPy/RadonPy), which is the first open-source Python library for fully automated calculation, for a comprehensive set of polymer properties, using all-atom classical MD simulations. For a given polymer repeating unit with its chemical structure, the entire process of the MD simulation can be performed fully automatically, including molecular modeling, equilibrium and nonequilibrium MD simulations, automatic determination of the completion of equilibration, scheduling of restarts in case of failure to converge, and property calculations in the post-process step. In this first release, the library comprises the calculation of 15 properties, such as the thermal conductivity, density, specific heat capacity, thermal expansion coefficient, and refractive index, in the amorphous state. In this study, we calculated 15 properties for more than 1,000 unique amorphous polymers. These calculated properties were systematically validated with respect to experimental values obtained from PoLyInfo. In particular, the focus here is on the thermal conductivity of polymers, which will be an important performance metric for designing polymeric materials used as insulating resins, molding resins, adhesives, and coating agents for mobile devices, given the increase in heat generation brought on by miniaturization and performance improvement of mobile devices. During the high-throughput data production, we computationally identified eight amorphous polymers with extremely high thermal conductivities, exceeding 0.4 W/m·K, including six polymers with unreported thermal conductivities. These polymers exhibited a high density of hydrogen bonding units or rigid, linear backbones. In addition, a decomposition analysis of the heat conduction, which is implemented in RadonPy, revealed the underlying mechanisms that yield such a high thermal conductivity: heat transfer via hydrogen bonds and dipole–dipole interactions between polymer chains having hydrogen bonding units or via covalent bonds of polymer backbones with high rigidity and linearity.

## 2. METHODS

The input parameter set for RadonPy comprises a simplified molecular input line entry system (SMILES)[23] string with two asterisks representing the connecting points of a repeating unit, the polymerization degree, the number of polymer chains in a simulation cell, and temperature. Subsequently, the following processes are fully automated (Figure 1): the conformation search for the repeating unit, calculation of the electronic properties, such as the atomic charge and dipole polarizability, based on the density functional theory (DFT), generation of initial configurations of polymer chains based on the self-avoiding random walk, assignment of the force field parameters, creation of a simulation cell such as an isotropic amorphous cell, MD simulation to equilibrate the system, determination of whether to reach equilibrium, execution of nonequilibrium MD simulation (NEMD) for thermal conductivity calculation, and calculation of various physical property values. In this study, we performed the automated calculation five times independently with different initial configurations for each polymer. RadonPy is mainly designed to run on a supercomputer; multiple polymers are calculated independently in parallel using many computation nodes in a supercomputer. The DFT and MD calculations were performed using the Psi4[24] and Large-scale Atomic/Molecular Massively Parallel Simulator (LAMMPS)[25], respectively, through the RadonPy interface. In the following, each step is described in detail.

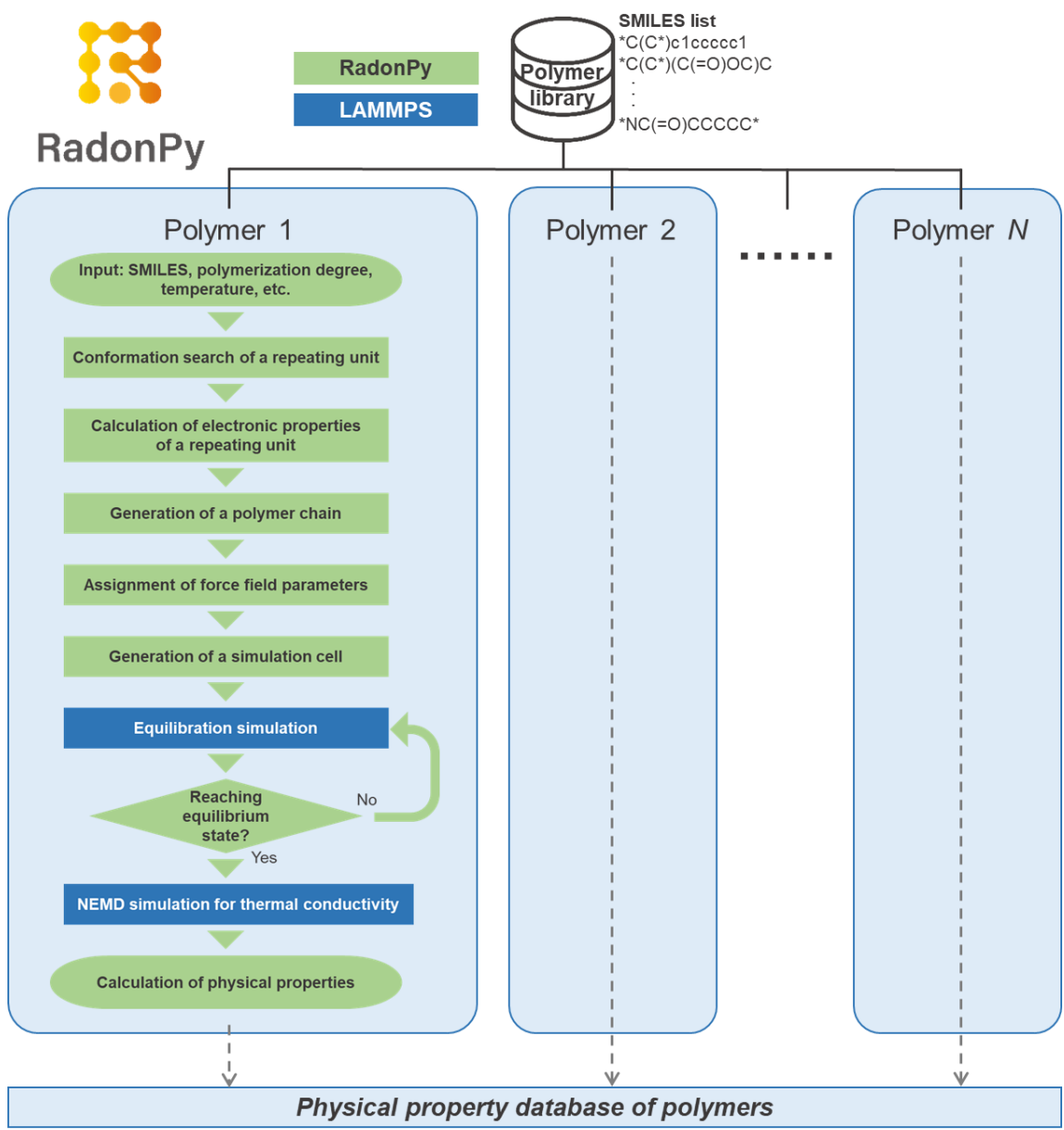

**Figure 1.** Flowchart of the automated MD calculation workflow for polymer properties using RadonPy.

*2.1. Model construction protocol*

*2.1.1. Conformation search of a repeating unit*

For a given SMILES string of a polymer repeating unit, 3D atomic coordinates of up to 1,000 different molecular conformations were generated using the ETKDG version 2 method[26–28] implemented in the Python library RDKit[29]. The SMILES string has two asterisk symbols for representing two attachment points of the repeating unit. These symbols were capped with hydrogen atoms. The potential energy of each conformation of a repeating unit was evaluated using the molecular mechanics calculation with the general Amber force field version 2 (GAFF2)[30,31] after the geometry optimization. Subsequently, the optimized conformers were clustered by performing the Butina clustering[32] based on the torsion fingerprint deviation.[33] The most stable four conformations were further optimized by performing DFT calculations with the ωB97M-D3BJ functional[34,35] combined with the 6-31G(d,p)[36,37] basis set. The most stable conformation was determined based on the DFT total energies.

*2.1.2 Calculation of electronic property of a repeating unit*

The atomic charges of a repeating unit were calculated using the restrained electrostatic potential (RESP) charge model[38] with a single-point calculation of the Hartree–Fock method[39] combined with the 6-31G(d) basis set on the optimized geometry of the most stable conformation. The total energy, the highest occupied molecular orbital (HOMO) energy level, the lowest unoccupied molecular orbital (LUMO) level, and the dipole moment were calculated with the single-point calculation using the ωB97M-D3BJ functional combined with the 6-311G(d,p) basis set[34,35,40–42] for H, C, N, O, F, P, S, Cl, and Br atoms and with the LanL2DZ basis set[43] for I atom. In addition, the dipole polarizability tensor was obtained by applying the finite field method under an electric field of $1.0 \times 10^{-4}$ a.u. using the ωB97M-D3BJ functional combined with the 6-311+G(2d,p) basis set[34,35,40–42,44,45] for H, C, N, O, F, P, S, and Cl atoms, with the 6-311G(d,p) for Br atom, and with the LanL2DZ basis set for I atom. The reason for using the 6-311+G(2d,p) basis set is that a basis set including double polarization function and diffuse function is required for appropriate polarizability calculations.[46] The isotropic dipole polarizability was defined as the mean of the diagonal values of the dipole polarizability tensor.

*2.1.3. Generation of polymer chains*

A polymer chain was constructed by connecting a repeating unit with the self-avoiding random walk algorithm. To prevent unintended chiral inversions and cis/trans conversions due to a large strain structure in the polymer chain growth, the bond between the head and capped atoms in a growing polymer chain and the bond of the tail and capped atoms in the next repeating unit were arranged to be coaxial and anti-parallel, the two capped atoms were deleted, and a new bond between the head and tail atoms was created. The length of the new bond was 1.5 Å, and the dihedral angle around the new bond was randomized in the range of −180° to +180° during the self-avoiding step. In this study, polymer chains were created to include approximately 1,000 atoms; thus, the degree of polymerization varies across polymers. By taking the number of atoms at the same level for different polymers, the molecular weights were controlled to be almost the same, and the dependency of the physical properties on the molecular weights was ignored here. The tacticity of a polymer chain could also be controlled in this process using RadonPy. In this study, all the polymers were generated as atactic polymers.

*2.1.4. Assignment of force field parameters*

The GAFF2 force field is expressed as follows:[30]

$$E_{MM} = \sum_{bonds} K_b(r - r_0)^2 + \sum_{angles} K_a(\theta - \theta_0)^2 + \sum_{dihedrals} K_d[1 + cos(n_d\varphi - \delta)] \quad (1)$$

$$+ \sum_{impropers} K_i(\chi - \chi_0)^2 + \sum_{i,j} \frac{q_i q_j}{4\pi\varepsilon_0 r_{ij}} + \sum_{i,j} 4\varepsilon_{ij}\left[\left(\frac{\sigma_{ij}}{r_{ij}}\right)^{12} - \left(\frac{\sigma_{ij}}{r_{ij}}\right)^{6}\right]$$

where $r$, $\theta$, $\varphi$, $\chi$, and $r_{ij}$ are the bond length, bond angle, dihedral angle, improper angle, and distance between atoms i and j, respectively; $K_b$, $K_a$, $K_d$, and $K_i$ denote the force constants of the bond, bond angle, dihedral angle, and improper angle, respectively; $r_0$, $\theta_0$, and $\chi_0$ are the equilibration structural parameters of the bond, bond angle, and improper angle, respectively; $n_d$ is the multiplicity, and $\delta$ is the phase angle for the torsional angle parameters; and $q_i$ and $q_j$ are the atomic charges of atoms i and j, and $\varepsilon_0$ is the dielectric constant of vacuum; $\varepsilon_{ij}$ and $\sigma_{ij}$ are the Lennard–Jones parameters determining the depth of the energy potential and equilibrium distance, respectively. The GAFF2 parameters of $K_b$ and $K_a$ were developed based on the force constant of the *ab initio* calculation and fine-tuned to reproduce the experimentally observed vibrational properties.[31] The parameter set was suitable for thermal conductivity calculations because the reproducibility of the vibrational properties was considered. The modified parameters for fluorocarbon developed by Träg and Zahn[47] were used for fluorocarbon polymers. The GAFF2 parameters were automatically assigned to each polymer chain in RadonPy. If the pre-defined parameter set lacked the bond angle parameters of $K_a$ and $\theta_0$ for a certain atom group, these parameter values were empirically estimated in the same manner as GAFF2.

*2.1.5. Generation of a simulation cell*

A simulation cell containing amorphous polymers was constructed by randomly arranging and rotating 10 polymer chains such that they did not overlap with each other, resulting in an amorphous cell having approximately 10,000 atoms. Initially, the density of the amorphous cell was set to 0.05 g/cm$^3$ and was then increased by conducting a packing simulation as described below.

*2.2. MD simulation protocol*

*2.2.1 Packing simulation*

The initial structure of the generated amorphous cell had a very low density. A packing simulation was performed to increase the density of the amorphous polymers to an appropriate value for subsequent calculations. A 1ns *NVT* simulation with a Nosé−Hoover thermostat was performed while the temperature was increased from 300 K to 700 K; in the next 1 ns *NVT* simulation, the calculation cell was isotropically reduced to a density of 0.8 g/cm$^3$ at 700 K. In this packing simulation, to prevent the self-aggregation of a polymer chain by intramolecular interactions leading to a globule-like structure, the Coulomb interaction was turned off, and the cutoff of the Lennard–Jones potential was set to 3.0 Å. Under this condition, the polymer chains remain random coil structures and could not pass through each other. Thus, the polymer chains were entangled in the final structure of the packing simulation. The time step was set to 1 fs, and all the bonds and angles, including those of the hydrogen atoms, were constrained by the SHAKE algorithm[48] in this packing simulation.

*2.2.2 Equilibration simulation*

The amorphous polymers after the packing simulation were equilibrated by the 21-steps compression/decompression equilibration protocol[49] proposed by Larsen and co-workers. In this protocol, a temperature rise to 600 K and a drop to 300 K were repeated for approximately 1.5 ns while the system was compressed to 50,000 atm and then decompressed to 1 atm by combining the *NVT* and *NpT* simulations with a Nosé−Hoover thermostat and a barostat. After the 21-steps equilibration, *NpT* simulations were run for more than 5 ns at 300 K and 1 atm until equilibrium was achieved. The achievement of the equilibrium was checked each 5 ns after the 21-steps equilibration. In this study, the equilibrium state was defined as being reached when the following conditions were met: the relative standard deviations (RSD) of the total, kinetic, bonding, bond angle, dihedral, van der Waals (vdW), and long-range coulomb energy fluctuations were less than 0.05, 0.05, 0.1, 0.1, 0.2, 0.2, and 0.1%, respectively, and the RSDs of the density and radius of gyration fluctuations were less than 0.1 and 1%, respectively. In this study, calculations that did not achieve equilibrium after 50 ns of equilibration calculations were treated as failures. The time step was set to 1 fs, and the SHAKE constraint[48] was applied to all the bonds and angles including those of the hydrogen atoms in this equilibration simulation. The twin-range cutoff method[50] was used for nonbonded interactions with a short cutoff of 8 Å and a long cutoff of 12 Å. The long-range Coulomb interaction was treated using the particle-particle particle-mesh (PPPM) method.[51] When the nematic order parameter decreased below 0.1, it was judged that the amorphous structure was appropriately generated; otherwise, it was treated as a failed calculation and removed from the data.

*2.2.3 NEMD simulation for thermal conductivity calculation*

To calculate the thermal conductivity, we performed the reverse NEMD simulation[52] proposed by Müller-Plathe. The simulation box of the reverse NEMD was constructed by triplication of an equilibrated amorphous cell in the *x*-axis direction. The reverse NEMD simulation involved dividing the simulation box into *N* slabs along the direction of the heat flux, which was generated in the system with temperature gradients induced by exchanging the velocity between the coldest atom in slab *N*/2 and the hottest atom in slab 0, as shown in Figure 2. To prevent the occurrence of temperature shifts due to cell replication, the preheating step with *NVT* ensemble was run for 2 ps at 300 K. Subsequently, the reverse NEMD with *NVE* ensemble was run for 1 ns. The number of slabs was set to 20, and the frequency of velocity swapping was set to 200 fs. The time step was set to 0.2 fs, and the SHAKE constraint was not applied in the reverse NEMD simulation. The twin-range cutoff method was used for nonbonded interactions with a short cutoff of 8 Å and a long cutoff of 12 Å. The long-range Coulomb interaction was treated using the PPPM method. As a validation of the adequacy of the reverse NEMD calculation, RadonPy confirmed a linearity in the temperature gradient. A calculation result with a poor linearity in the temperature gradient ($R^2$ less than 0.95) was treated as a failure and removed from the data.

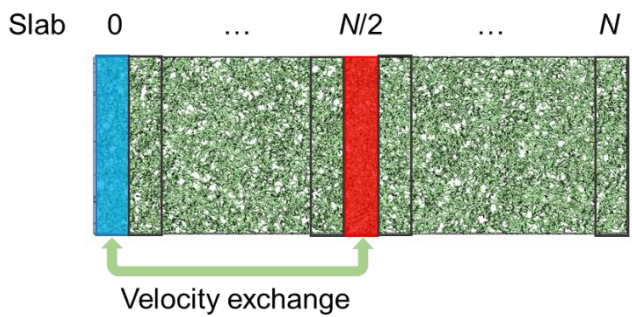

**Figure 2.** Schematic representation of the simulation box for reverse nonequilibrium molecular dynamics.

A thermal conductivity decomposition analysis was performed for 100 ps. For the Irving-Kirkwood equation[53] modified by Torii and co-workers,[54] the energy flux can be expressed as follows:

$$J_u = \frac{1}{V} \left\{ \sum_{i \in V} e_i v_{i,u} + \sum_{i \in V} (S_i v_i)_u \right\} \quad (2)$$

where $J_u$ is the energy flux along the direction of unit vectors $u$, $V$ is the volume, $e_i$ is the per-atom potential and kinetic energy, $v_{i,u}$ is the velocity of the atom, $S_i$ is the per-atom stress tensor, and i is the index of atoms. The first and second terms represent the contribution to the energy flux via convection and interatomic interactions, respectively. The second term can be further divided into each component of the interactions. The component (a, b) of the stress tensor can be written as[55–57]

$$S_{ab} = \sum_{n=1}^{N_p} r_{i0,a} F_{i,b} + \sum_{n=1}^{N_b} r_{i0,a} F_{i,b} + \sum_{n=1}^{N_a} r_{i0,a} F_{i,b} + \sum_{n=1}^{N_d} r_{i0,a} F_{i,b} + \sum_{n=1}^{N_i} r_{i0,a} F_{i,b} \quad (3)$$
$$+ \text{Kspace}(r_{i,a}, F_{i,b})$$

The first to fifth terms denote the pairwise, bond, angle, dihedral, and improper contributions, respectively, where $F_i$ denotes the force acting on atom i due to the interaction, $r_{i0}$ denotes the relative position of atom i to the geometric center of its interacting atoms, and $N_p$, $N_b$, $N_a$, $N_d$, and $N_i$ are the numbers of atom pairs, bonds, bond angles, dihedral angles, and improper angles, respectively. The sixth term is the K-space contribution from the long-range Coulombic interactions. The partial thermal conductivity $\lambda_{\text{partial}}$ is given by

$$\lambda_{\text{partial}} = \frac{J_{\text{partial}}}{J_{\text{total}}} \lambda_{\text{total}} \quad (4)$$

where $J_{\text{total}}$ is the total heat flux calculated by Eq 2, $J_{\text{partial}}$ is the partial heat flux subdivided by each term of Eqs 2 and 3, and $\lambda_{\text{total}}$ is the total thermal conductivity calculated by the reverse NEMD.

*2.3. Calculation of physical properties*

The density in a *NpT* simulation was computed using the mass *m* and volume *V* of the system as follows:

$$\rho = \frac{m}{\langle V \rangle} \quad (5)$$

where the angular brackets $\langle \cdot \rangle$ represent time averaging.

The radius of gyration *Rg* was calculated using the following equation:

$$Rg = \sqrt{\frac{1}{N}\sum_{k=1}^{N}(\mathbf{r}_k - \mathbf{r}_{mean})^2} \quad (6)$$

where $\mathbf{r}_k$ is the position of a repeating unit k, and $\mathbf{r}_{mean}$ denotes the mean position of the repeating units in a polymer chain.

The specific heat capacity at constant pressure $C_p$ was calculated from the fluctuations in the enthalpy *H*:[58]

$$C_P = \frac{\langle \delta H^2 \rangle}{k_B T^2 m} \quad (7)$$

where $k_B$ is the Boltzmann constant, and *T* is the temperature. The enthalpy was calculated using the constant pressure of 1 atm because the calculated pressure value in the *NpT* simulations has a significant fluctuation, leading to inaccurate $C_P$ calculation.

The isothermal compressibility $\beta_T$ and isothermal bulk modulus $K_T$ were calculated from the fluctuations in the volume *V*:[58]

$$\beta_T = \frac{\langle \delta V^2 \rangle}{k_B T \langle V \rangle} \quad (8)$$

$$K_T = \frac{1}{\beta_T} \quad (9)$$

The volumetric thermal expansion coefficient $\alpha_P$ was calculated from the covariance of the volume *V* and enthalpy *H*:[58]

$$\alpha_P = \frac{\langle \delta V \delta H \rangle}{k_B T^2 \langle V \rangle} \quad (10)$$

Here, the enthalpy was calculated at a constant pressure of 1 atm. The linear thermal expansion coefficient $\alpha_{P,l}$ in the isotropic systems was calculated using the following equation:[58]

$$\alpha_{P,l} = \frac{\alpha_P}{3} \tag{11}$$

The specific heat capacity at constant volume $C_V$ was calculated from the following equation, associated with $C_P$, $\alpha_P$, and $\beta_T$:[58]

$$C_V = C_P - \frac{\alpha_P^2 T \langle V \rangle}{\beta_T m} \tag{12}$$

The isentropic compressibility $\beta_S$ and isentropic bulk modulus $K_S$ were calculated using the following equations:

$$\beta_S = \beta_T \frac{C_V}{C_P} \tag{13}$$

$$K_S = \frac{1}{\beta_S} \tag{14}$$

The self-diffusion coefficient was calculated using the Einstein equation:[58]

$$D = \lim_{t \to \infty} \frac{1}{6t} \langle |\mathbf{r}(t + t_0) - \mathbf{r}(t_0)|^2 \rangle \tag{15}$$

where $t$ is the time, and $\mathbf{r}$ denotes the atomic position at the time.

The refractive index $n$ was obtained from the Lorentz–Lorenz equation:

$$\frac{n^2 - 1}{n^2 + 2} = \frac{4\pi}{3} \frac{\rho}{M} \alpha_{polar} \tag{16}$$

where $\alpha_{polar}$ is the isotropic dipole polarizability of a repeating unit computed from the DFT calculation, and $M$ is the molecular weight of a repeating unit.

The static dielectric constant $\varepsilon(0)$ was calculated using the equation:[59]

$$\varepsilon(0) = \frac{\langle \mathbf{\mu}^2 \rangle - \langle \mathbf{\mu} \rangle^2}{3\varepsilon_0 k_B T \langle V \rangle} + \varepsilon_{el} \tag{17}$$

where $\mathbf{\mu}$ is the dipole moment of the system, $\varepsilon_0$ is the dielectric constant of vacuum, and $\varepsilon_{el}$ is the contribution of the electronic polarization in the dielectric constant, which is evaluated as the square of the refractive index $n^2$.

The nematic order parameter was calculated as the highest eigenvalue of the second rank ordering tensor[58] $Q_{\alpha\beta}$, following equation:

$$Q_{\alpha\beta} = \frac{1}{N}\sum_{i=1}^{N}\frac{1}{2}(3u_{i\alpha}u_{i\beta} - \delta_{\alpha\beta}) \tag{18}$$

where $u_{i\alpha}$ and $u_{i\beta}$ ($\alpha, \beta = x, y,$ or $z$) are the unit vectors of the molecular axis of a repeating unit i, $\delta_{\alpha\beta}$ is the Kronecker delta, and $N$ is the number of repeating units. The molecular axis of each repeating unit is defined as the long axis found from the inertia tensor. The nematic order parameter takes on a value between 0 for an isotropic structure and 1 for a completely ordered structure.

The thermal conductivity $\lambda$ was calculated according to Fourier's law:

$$\lambda = \frac{J_Q}{(\partial T/\partial x)} = \frac{\Delta E}{2A\,\Delta t\,(\partial T/\partial x)} \tag{19}$$

where $J_Q$ is the heat flux, and $\partial T/\partial x$ is the temperature gradient of the NEMD simulation. The heat flux $J_Q$ can be calculated from the exchanged energy obtained using the Müller-Plathe algorithm $\Delta E$, the cross-sectional area in the heat flux direction $A$, and the simulation time $\Delta t$. The thermal diffusivity $\kappa$ was obtained from the calculated thermal conductivity $\lambda$, density $\rho$, and heat capacity $C_P$:

$$\kappa = \frac{\lambda}{\rho C_P} \tag{20}$$

*2.4. Dataset*

The PoLyInfo database contains 15,335 homopolymers, which have only organic 10 element species, H, C, N, O, F, P, S, Cl, Br, and I. Among these, we selected 1,138 unique homopolymers as the calculation target in this study, for which as many experimental properties as possible were recorded. The selected polymer set was composed of a wide variety of polymer backbones, such as polystyrenes, polyvinyl, polyacrylates, polyamides, polycarbonates, polyurethanes, and polyimides. The validation data of the density, thermal conductivity, refractive index, specific heat capacity $C_P$, linear expansion coefficient, and volumetric expansion coefficient were collected from PoLyInfo. The data used were limited to homopolymers and those meeting the following conditions: their material type was labeled as one of neat resin, samples contained no additives, fillers, and dopants, the measured temperature was in the range of 273–323 K, the postforming state was amorphous or unidentified, and the topology of the polymers was linear or unidentified.

## 3. RESULTS AND DISCUSSION

*3.1. Distribution of calculated polymers in chemical space*

The automated MD calculations were conducted for the 1,138 homopolymers selected from the PoLyInfo database. Of the five independent calculations, the automatic calculations succeeded at least once for 1,070 polymers, more than thrice for 1,001 polymers, and in all the five cases for 759 polymers. The calculated dataset is provided in the Supporting Information. The failed calculations were classified into four cases: the structural optimization of the DFT calculation did not converge, the MD simulation did not reach equilibrium, the system was partially oriented (nematic order parameter > 0.1), and the temperature gradient in the NEMD calculation did not become linear.

To investigate the distribution of the backbones of the 1,070 calculated polymers over the 15,335 polymers in PoLyInfo, their chemical structures were visualized onto a 2D space using the uniform manifold approximation and projection (UMAP)[60]. The chemical structure of each polymer was transformed into a 2,048 bit vector with an extended connectivity fingerprint with a radius of 3 atoms (ECFP6)[61]. To consider the repeating structure of polymers, the ECFP6 descriptor was constructed after generating the macrocyclic oligomer with 10-mer of the repeating unit. The UMAP with the Hamming distance was used to create the 2D representation of the 15,335 fingerprinted polymers, as shown in Figure 3(a), in which its subset corresponding to the 1,070 polymers successfully calculated at least once is shown in Figure 3(b). The plot colors indicate the 21 classes of the polymer backbones according to the PoLyInfo database. The two distributions exhibited similar patterns in the UMAP plot, confirming no significant selection bias in the calculated polymers. In addition, the calculated polymers were selected to cover the 20 classes except for the class of others.

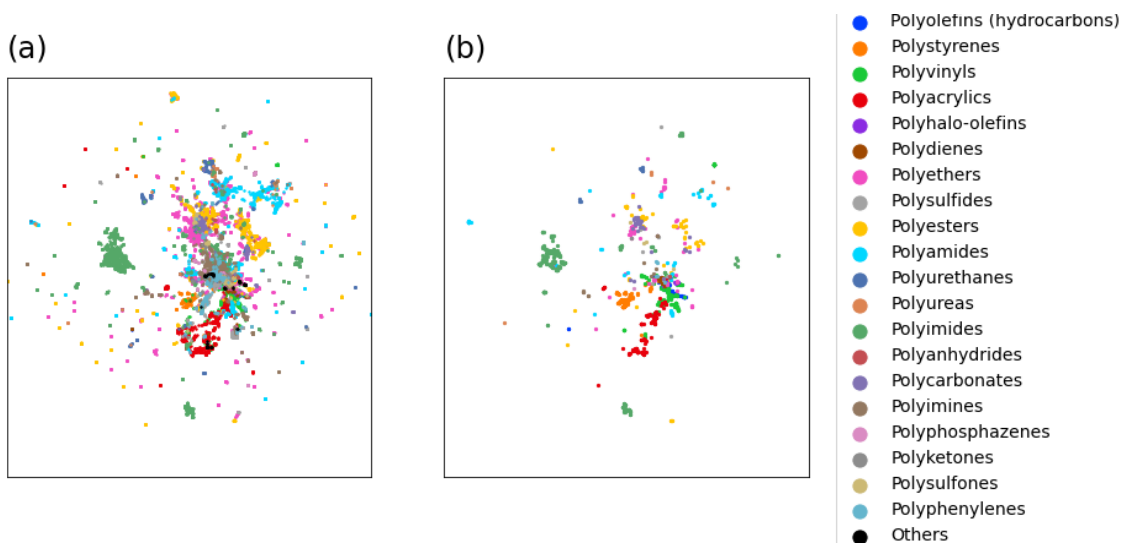

**Figure 3.** UMAP plot visualizing the distribution of the polymer backbones of (a) 15,335 homopolymers in PoLyInfo and (b) 1,070 homopolymers calculated in this study. The 21 classes of the polymer backbones are color-coded according to the definition of PoLyInfo.

*3.2. Validation of the calculated physical properties*

To evaluate the performance of the automated MD pipeline, the calculated properties were systematically compared with the experimental values from PoLyInfo (Figure 4). In the validation process, we used the 1,001 polymers for which the automatic calculation was successfully completed at least thrice out of the five independent trials. The calculated density well reproduced the experimental values ($R^2 = 0.890$), albeit with a slight underestimation, as the slope of the fitted straight line in the parity plot was equal to 0.805 in Figure 4(a). The standard deviation (SD) of the calculated values in the five independent trials was low. The slight underestimation may be explained that, since the polymerization degree in our MD calculations was smaller than that in the experimental values from PoLyInfo, a mobility of polymer chains in the MD calculations was larger than that in the experimental values, resulting in a slight overestimation of free volume in the MD calculations. As reported in a previous study, the MD calculation of the density of the organic molecule liquids using the GAFF2 force field is often poorly performed in high-density regions.[62] On the other hand, in our calculation, such a discrepancy never occurred in the high-density regions. This is because of the use of the modified force field parameters developed by Träg and Zahn[47] for fluorocarbon polymers.

The calculated thermal conductivities also showed good agreement with the experimental values in PoLyInfo ($R^2 = 0.490$), as shown in Figure 4(b). However, the correlation was not so high. This could be because the experimental values of the thermal conductivity involve fluctuations due to differences in the measurement methods and temperature dependence. In addition, as the level of the thermal conductivity increases, the fluctuation in the calculated values within the independent trials increases significantly (Figure S1). Thus, for polymers with potentially high thermal conductivity, the number of independent trials of MD calculations should be increased to improve the accuracy.

The calculated refractive index well reproduced the PoLyInfo dataset ($R^2 = 0.809$) with a trivial underestimation where the slope of the fitted straight line in the parity plot was equal to 0.839 (Figure 4(c)). The slight underestimation can be attributed to the slight underestimation of density because the refractive index is the function of density. The variation in the MD simulations was quite small. It can be concluded that a sufficiently high prediction accuracy was obtained for the refractive index.

Figure 4(d) shows the correlation of $C_P$ between the calculated and experimental values ($R^2 = 0.602$). The calculated $C_P$ showed an evident overestimation as the fitted slope in the parity plot was 1.430. This observation is inevitable in the classical MD because classical MD calculations do not include quantum effects: the vibrational energy in a classical harmonic oscillator is significantly higher than that in a quantum harmonic oscillator at the same frequency. Fortunately, since the observed correlation is relatively clear, it would not be difficult to correct the systematic bias by applying e.g., transfer learning or multi-fidelity learning.

The linear and volume expansion coefficients showed weak correlations ($R^2 = 0.178$ and $0.217$) between the calculated and experimental values (Figures 4(e) and (f)). The variations in the linear and volume expansion coefficients within the same polymer were significant both experimentally and computationally. Previous studies also reported that MD-calculated values of the volumetric expansion coefficients in molecular organic liquids are highly variable and less reproducible with respect to the experimental values.[62] Possibly, the timescale and simulation cell size of the MD simulations should be sufficiently large for an accurate simulation.

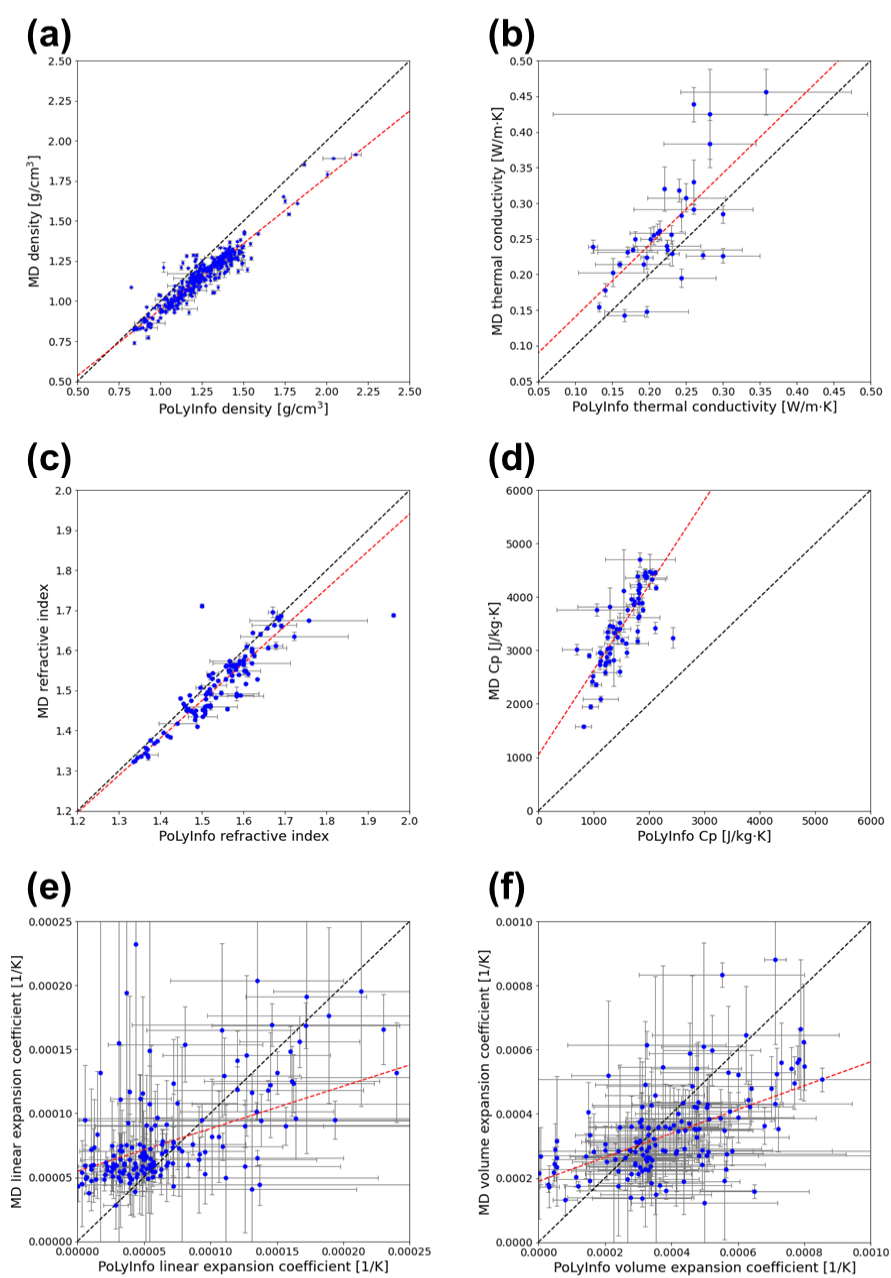

**Figure 4.** Comparison between the MD-calculated properties of various amorphous polymers (vertical axis) and their experimental values in PoLyInfo (horizontal axis) for (a) density (N = 382), (b) thermal conductivity (N = 34), (c) refractive index (N = 107), (d) specific heat capacity (N = 66), (e) linear expansion coefficient (N = 165), and (f) volume expansion coefficient (N = 144). The error bar indicates the standard deviation of the calculated or measured properties within the same polymer. The dashed black line indicates the y = x line. The red line is the regression line fitted to the calculated and experimental values.

*3.3. Data distribution*

The marginal distributions of the six properties for the calculated 1,070 unique amorphous polymers are presented in the diagonal panels of Figure 5, and their statistics are summarized in Table 1. The calculated thermal conductivities were distributed between 0.082 and 0.619 W/m·K, with their mean being 0.240 W/m·K. The thermal conductivity of the unoriented polymers in the amorphous states is known to be typically less than 0.3 W/m·K. On the other hand, few of the calculated polymers exhibited exceedingly high thermal conductivities. However, as mentioned above, in the high-thermal-conductivity regions, the fluctuations in the MD-calculated properties became significant. Thus, we narrowed down to eight highly reliable polymers, as shown in Figure 6, with small variation in the repeated calculations (SD < 0.05 W/m·K). For polyethylene (**PI1**) and poly(vinyl alcohol) (**PI241**), the experimental values of thermal conductivity can be found in the PoLyInfo. The calculated thermal conductivity of **PI1** was 0.456 W/m·K, which is consistent with the reported values (0.39–0.53 W/m·K) of polyethylene neat resin in PoLyInfo. The calculated thermal conductivity of **PI241** was 0.439 W/m·K, which is overestimated compare to the reported values (0.31 W/m·K) of poly(vinyl alcohol) neat resin in PoLyInfo. The experimental values of thermal conductivity of the other six polymers were not recorded in the PoLyInfo. Apart from polyethylene (**PI1**), the structural features of these polymers fall into three types: (1) polymers with a high density of hydrogen bonding units (**PI241** and **PI305**), (2) aromatic polyamides with rigid, linear backbones (**PI626**), and aromatic polyimides (**PI687**, **PI711**, **PI715**, and **PI1093**).

In addition, the calculated values of the density and refractive index sufficiently correlated with the experimental values; therefore, we investigated the distributions of these properties from a quantitative viewpoint. The calculated density values were distributed between 0.742 and 1.914 g/cm$^3$ with their mean being 1.133 g/cm$^3$. Twelve amorphous polymers were identified as having a high-density state: >1.75 g/cm$^3$. These polymers were found to contain rich halogen atoms (Figure S2). The calculated values of the refractive index ranged from 1.274 to 1.839 with their mean equal to 1.550. Nine polymers were identified as high-refractive-index polymers in amorphous states, with their refractive index being greater than 1.75. These polymers had large π-conjugated backbones (Figure S3), indicating that the calculated high refractive index originated from the high polarizability of the large π-conjugation.

An observation of the joint distribution of the multiple properties, as shown in the off-diagonal panels of Figure 5, provides hypothetical insights into the hidden dependency of the multiple properties, and the existence and location of the Pareto frontiers with the chemical features of the constituent polymers. The observed Pareto frontier of the specific heat capacity and thermal conductivity suggests the difficultly of achieving both high specific heat capacity and low thermal conductivity in amorphous polymers. The polymers distributed around the Pareto frontier were mainly polystyrenes and polyacrylates. On the other hand, no Pareto frontier was observed in the region of higher thermal conductivity. The joint distribution of the thermal conductivity and refractive index shows that there are still unexplored regions of amorphous polymers reaching lower thermal conductivity with higher refractive index and higher thermal conductivity with a lower refractive index. The specific heat capacity was inversely proportional to the density. This observation can be explained by the Dulong–Petit law.[63] The specific heat capacity is inversely proportional to the mean atomic weights (Figure S4) because the heat capacity of a mole is typically almost a constant in materials. On the other hand, the density is proportional to the mean atomic weights (Figure S5). In the joint distribution of the density and refractive index, their correlation was unclear. According to the Lorentz–Lorenz equation (Eq 16), the refractive index is described as a function of the density and polarizability. The observed distribution implies that, for polymers in amorphous states, the polarizability is dominant in determining the refractive index.

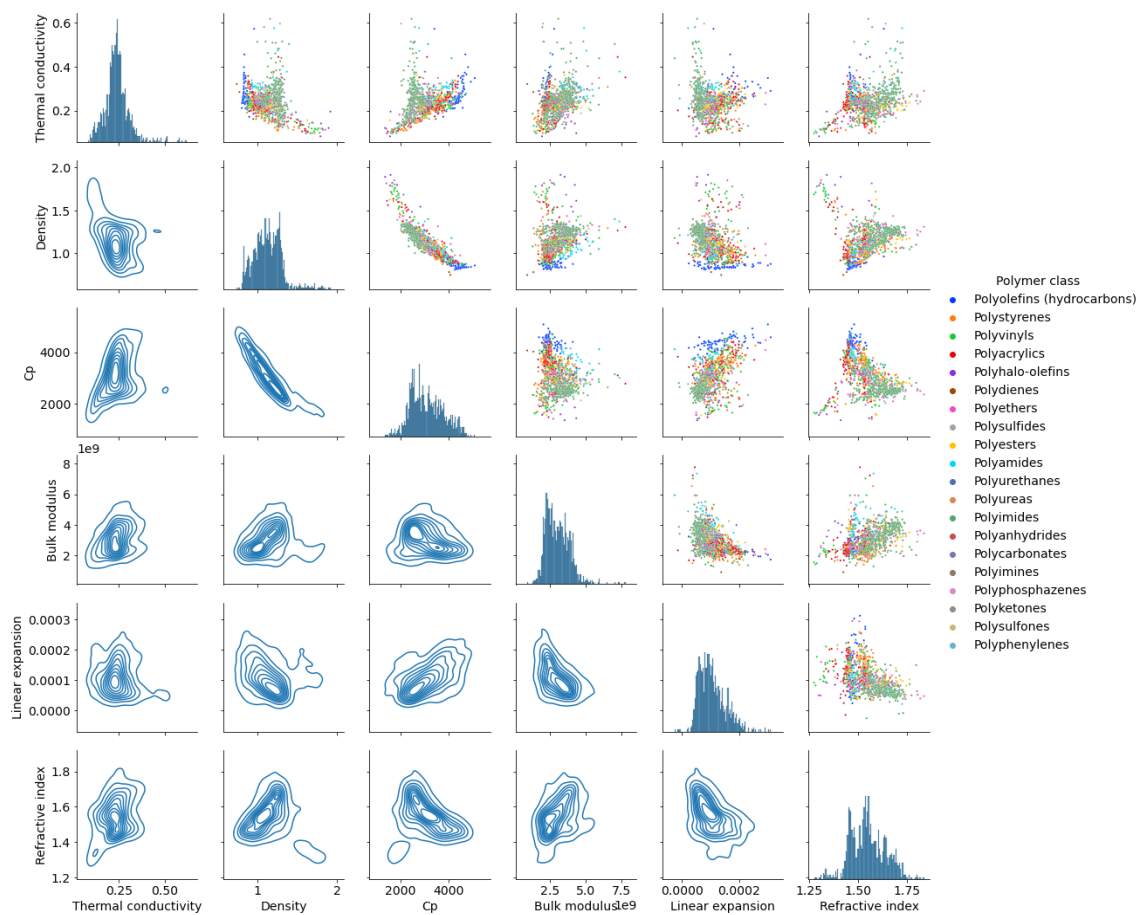

**Figure 5.** Joint distribution of the six properties calculated from the automatic MD simulation, including the thermal conductivity (W/m·K), density (g/cm$^3$), specific heat capacity $C_P$ (J/kg·K), bulk modulus (Pa), linear expansion coefficient (1/K), and refractive index. The diagonal panels represent the histograms of the individual property values. In the upper off-diagonal panels, a scatter plot of each pair of properties is displayed. The lower off-diagonal panels represent the kernel density estimation of the bivariate joint distributions, which is displayed with contours.

**Table 1.** Summary statistics of the calculated properties, including the mean, standard deviation (SD), minimum, and maximum.

| Properties | Number of polymers | Mean | SD | Minimum | Maximum |
|---|---|---|---|---|---|
| Thermal conductivity (W/m·K) | 1070 | 0.240 | $6.562\times10^{-2}$ | $8.220\times10^{-2}$ | 0.619 |
| Thermal diffusivity (m$^2$/s) | 1069 | $7.100\times10^{-8}$ | $2.014\times10^{-8}$ | $2.957\times10^{-8}$ | $2.273\times10^{-7}$ |
| Density (g/cm$^3$) | 1077 | 1.133 | 0.180 | 0.742 | 1.914 |
| Radius of gyration (Å) | 1077 | 20.59 | 8.149 | 10.37 | 85.68 |
| Self-diffusion coefficient (m$^2$/s) | 1076 | $6.747\times10^{-13}$ | $8.693\times10^{-13}$ | $8.939\times10^{-15}$ | $1.098\times10^{-11}$ |
| $C_P$ (J/kg·K) | 1076 | 3086 | 691.6 | 1345 | 4955 |
| $C_V$ (J/kg·K) | 1076 | 2993 | 644.3 | 1331 | 4579 |
| Compressibility (1/GPa) | 1076 | 0.360 | 0.107 | 0.129 | 1.299 |
| Isentropic compressibility (1/GPa) | 1076 | 0.349 | 0.102 | 0.128 | 1.255 |
| Bulk modulus (GPa) | 1076 | 3.062 | 0.835 | 0.921 | 7.766 |
| Isentropic bulk modulus (GPa) | 1076 | 3.144 | 0.842 | 0.935 | 7.862 |
| Linear expansion coefficient (1/K) | 1076 | $1.048\times10^{-4}$ | $4.611\times10^{-5}$ | $-2.598\times10^{-5}$ | $3.115\times10^{-4}$ |
| Volumetric expansion coefficient (1/K) | 1076 | $3.144\times10^{-4}$ | $1.383\times10^{-4}$ | $-7.794\times10^{-5}$ | $9.345\times10^{-4}$ |
| Static dielectric constant | 1075 | 4.866 | 10.535 | 1.674 | 130.8 |
| Refractive index | 1075 | 1.550 | $8.857 \times 10^{-2}$ | 1.274 | 1.839 |
| Properties of a repeating unit | | | | | |
| HOMO (eV) | 1077 | -9.205 | 0.918 | -7.458 | -12.647 |
| LUMO (eV) | 1077 | 1.016 | 1.416 | -5.997 | 3.188 |
| Dipole moment (Debye) | 1077 | 2.426 | 1.948 | $6.175 \times 10^{-7}$ | 12.14 |
| Dipole polarizability ($Å^3$) | 1077 | 35.45 | 25.73 | 3.842 | 139.6 |

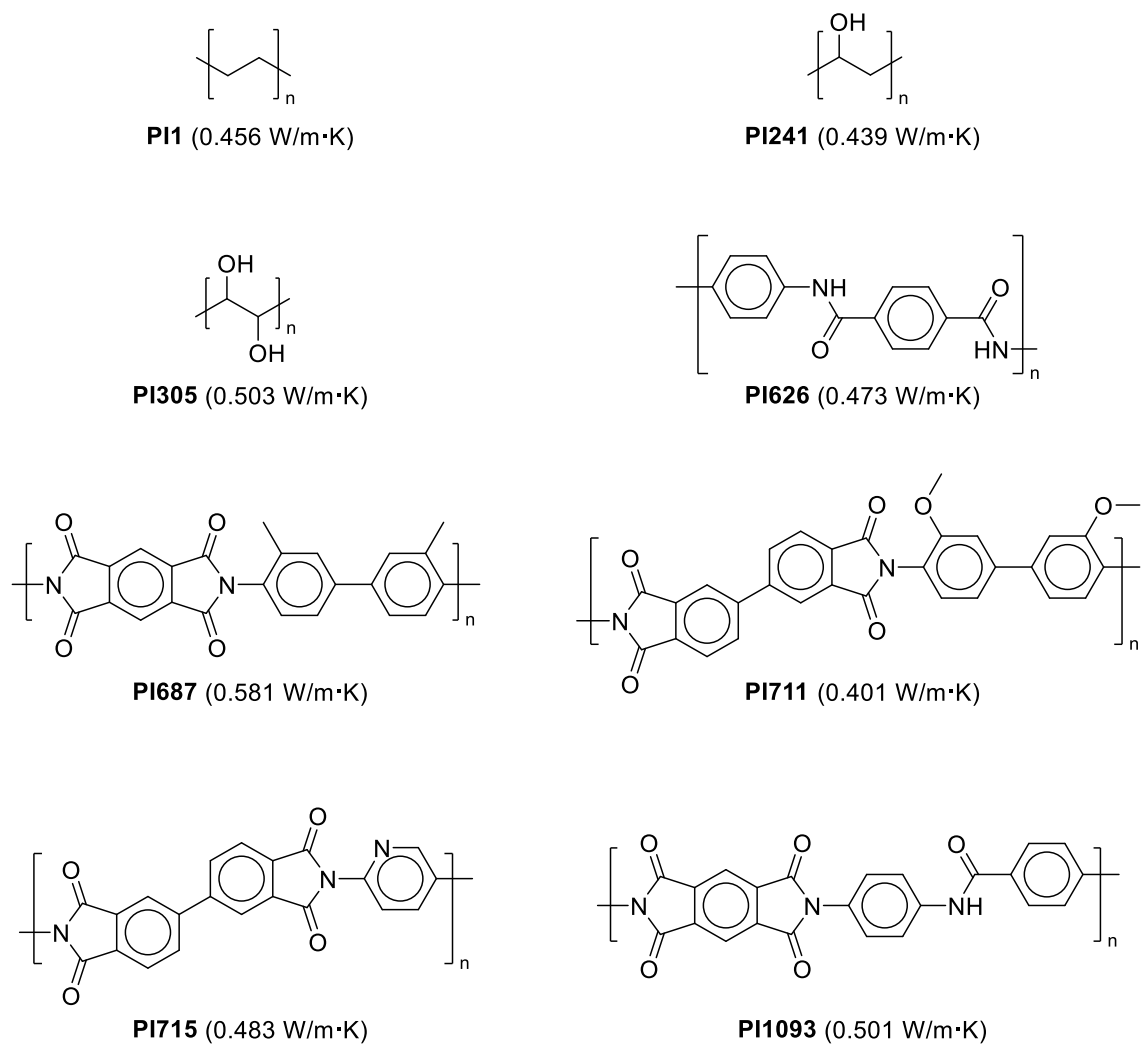

**Figure 6.** Repeating units of identified polymers exhibiting a high thermal conductivity in amorphous states. The compound identifier corresponds to the polymer ID in the calculated dataset provided in the Supporting Information.

*3.4. Decomposition analysis of thermal conductivity*

The decomposition analysis was performed to understand the mechanism of the eight polymers (Figure 6) that exhibited a high thermal conductivity. As shown in Figure 7, for each calculated thermal conductivity, the decomposition analysis quantified the contribution of the six components corresponding to convection, bond, angle, dihedral, improper, and nonbonded, where the nonbonded contribution represents the sum of the pairwise and K-space contributions described in Eq 3. Since the contribution of the improper term was negligible, it is shown as a dihedral term in Figure 7. Notably, the AMBER-type force field describes the dihedral potential as the sum of the dihedral term and nonbonded 1–4 interactions; thus, a part of the nonbonded contribution is essentially attributed to the dihedral contribution.[30]

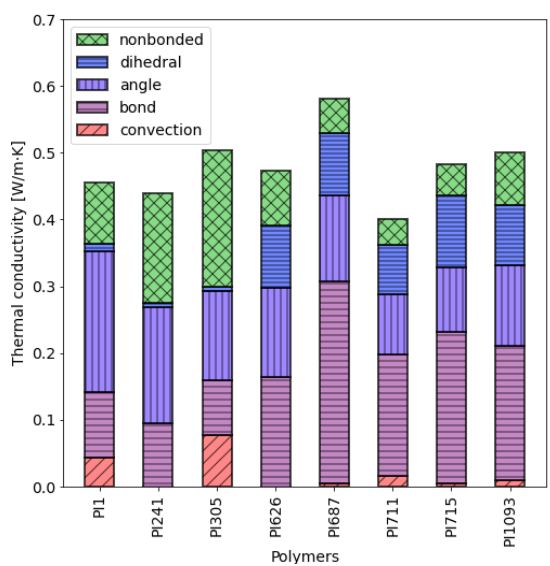

**Figure 7.** Contributions of convection and different types of interactions to the calculated high thermal conductivities of the eight polymers.

The calculated thermal conductivity of **PI1** (polyethylene) was 0.456 W/m·K. The high thermal conductivity of **PI1** was due to the significant contribution of bond bending.

The high thermal conductivities of **PI241** (polyvinyl alcohol) and **PI305** (poly(vinylene) carbonate) were largely due to the contributions of nonbonded interactions. The polymer chains of **PI241** and **PI305** contain highly condensed hydroxyl groups. This indicates that a high density of hydrogen bonding units provides large intermolecular interactions via the creation of hydrogen bonds and dipole–dipole interactions, resulting in a significant contribution of nonbonded interactions. Thus, the thermal conductivities of **PI241** and **PI305** are enhanced by the heat transfer via hydrogen bonds and dipole–dipole interactions.

In the aromatic polyamide **PI626** (poly-*p*-phenyleneterephthalamide a.k.a. Kevlar), the bond, angle, dihedral, and nonbonded interactions showed moderately large contributions. The results can be explained as follows: the backbone of **PI626** is relatively rigid, resulting in a significant contribution to the thermal conductivity through covalent bonds, and **PI626** can create the interaction of hydrogen bonds and dipole–dipole interactions with its amide groups, resulting in moderately high contributions through nonbonded interactions.

Thermally conductive behaviors in the aromatic polyimides **PI687**, **PI711**, **PI715**, and **PI1093** were largely due to the contributions of bond stretching. The **PI687** had a significantly large contribution of bond stretching. Aromatic polyimides have rigid backbones, particularly **PI687**, which has high rigidity and linearity. The results show that the rigid and linear characteristics of a polymer backbone can help enhance the thermal conductivity through the contribution of bond stretching. The **PI1093** is an aromatic polyimide containing an amide group. The contribution of nonbonded interactions of **PI1093** was the largest in the four identified aromatic polyimides. This suggests that polymers containing hydrogen bonding units and having rigid and linear backbones can help further increase the thermal conductivity in amorphous states.

Figure 8 shows the joint distribution of the total thermal conductivity with each quantified contribution. The correlations with the total thermal conductivity can be clearly observed in the bond, angle, dihedral, and nonbonded terms. On the other hand, the convection term did not correlate significantly with the thermal conductivity. In summary, thermally conductive amorphous polymers can be designed, in principle, by increasing the contributions of the bond, angle, dihedral, and nonbonded terms.

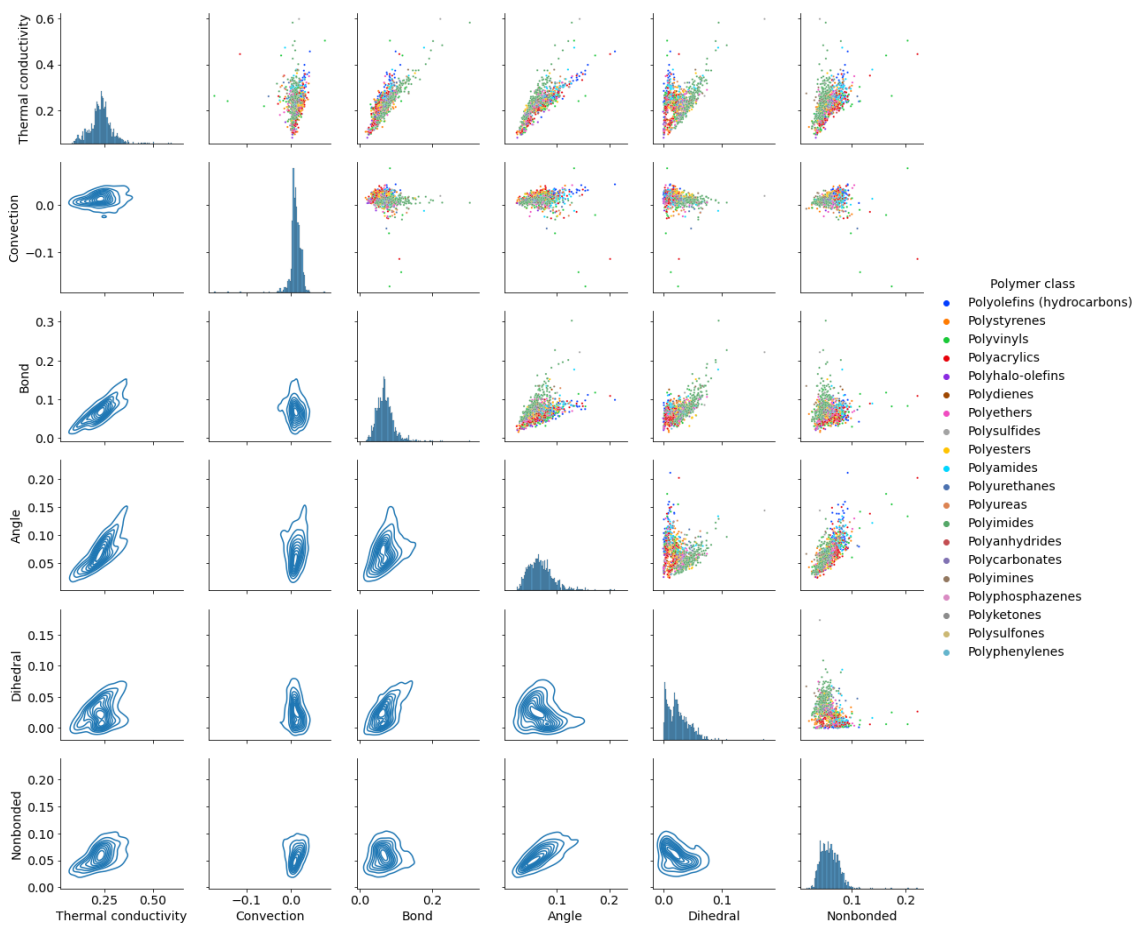

**Figure** 8. Distribution of the thermal conductivity (W/m·K) and categorization in terms of its contributions from convection, bond, angle, dihedral, and nonbonded terms. Diagonal panels represent the histograms of individual quantities. In the upper off-diagonal panels, the scatter plots of the six quantities are displayed. The lower off-diagonal panels represent their kernel density estimation, which is displayed with contours.

## 4. CONCLUSIONS

We presented RadonPy, which is the first open-source Python library to fully automate polymer property calculations using all-atom classical MD simulations. The high-throughput calculation using RadonPy was successfully performed for more than 1,000 unique amorphous polymers with a wide variety of thermophysical properties, such as the thermal conductivity, refractive index, density, and specific heat capacity $C_P$. For systems other than amorphous homopolymers, such as copolymers, blend polymers, and uniaxially oriented systems, as well as for other properties, automated calculation capabilities have already been implemented; however, no calculation protocols based on experimental data have been established. In RadonPy, automatic calculation protocols for various polymer properties can be implemented as an add-on feature. We will continue to promote the development of RadonPy.

In this study, the agreement between a total of six properties obtained from the high-throughput MD calculation and experimental values was comprehensively verified. As a result, the refractive index, density, and thermal conductivity successfully reproduced the experimental values quantitatively. The calculated values of the specific heat capacity were also highly correlated with the experimental values, although the classical MD calculation had a systematic bias due to its inability to represent quantum effects. For the linear and volumetric expansion coefficients, the correlation between the calculated and experimental values was weak due to large variations and uncertainties in both the calculations and experiments. There has been no previous work on a comprehensive validation of high-throughput MD simulations of polymer properties on such a scale. More rigorous comprehensive validation with experimental values, including other properties not discussed in this study, should be conducted to determine appropriate calculation conditions and protocols.

Compared with other material systems, polymer research has lagged in terms of constructing open databases available for data-driven research. The primary objective in the development of RadonPy was to use it to create a systematically designed polymer property database. In the early days of MI in inorganic chemistry, the development of an open database was strategically promoted. In particular, huge computational property databases constructed using high-throughput first-principles calculations drove the evolution and widespread applications of MI. Large-scale computational property data have historically proven to be an important resource in MI, and RadonPy was designed for the rapid production of large amounts of polymer property data using highly parallel computers such as supercomputers. In this study, more than 1,000 unique amorphous polymers were computed in approximately two months mainly using the world's fastest supercomputer, Fugaku. In the future, our growing data will significantly facilitate the evolution of polymer informatics, just like the first-principles computational database for inorganic crystals.

**ACKNOWLEDGMENTS**

The numerical calculations were conducted on the supercomputer Fugaku at the RIKEN Center for Computational Science, Kobe, Japan, on the supercomputer at the Research Center for Computational Science, Okazaki, Japan, on the supercomputer Ohtaka at the Supercomputer Center, the Institute for Solid State Physics, the University of Tokyo, Tokyo, Japan, on the supercomputer TSUBAME3.0 at the Tokyo Institute of Technology, Tokyo, Japan, and on the supercomputer ABCI at the National Institute of Advanced Industrial Science and Technology, Tsukuba, Japan. This work was supported by a JST CREST (Grant Number JPMJCR19I3 to J.M. and R.Y.), the MEXT as "Program for Promoting Researches on the Supercomputer Fugaku" (Project ID: hp210264 to R.Y.), the Grant-in-Aid for Scientific Research (A) from the Japan Society for the Promotion of Science (19H01132 to R. Y.), and the HPCI System Research Project (Project ID: hp210213 to Y.H.). The authors are grateful to Hiroki Sugisawa from Mitsubishi Chemical Corporation and the RadonPy consortium members for helpful discussions about molecular dynamics simulations. The authors are also grateful to the PoLyInfo development team at National Institute for Materials Science for providing benchmark data.

# Supporting Information

# RadonPy: Automated Physical Property Calculation using All-atom Classical Molecular Dynamics Simulations for Polymer Informatics

Yoshihiro Hayashi,[1,2]* Junichiro Shiomi,[1,2] Junko Morikawa,[1,3] Ryo Yoshida[1,4,5]*

[1] Data Science Center for Creative Design and Manufacturing, The Institute of Statistical Mathematics (ISM), Research Organization of Information and Systems, 10-3 Midori-cho, Tachikawa, Tokyo 190-8562, Japan

[2] Department of Mechanical Engineering, The University of Tokyo, 7-3-1 Hongo, Bunkyo, Tokyo 113-8656, Japan

[3] Department of Materials Science and Engineering, School of Materials and Chemical Technology, Tokyo Institute of Technology, 2-12-1-E4-6 Ookayama, Meguro-ku, Tokyo 152-8552, Japan

[4] Research and Services Division of Materials Data and Integrated System (MaDIS), National Institute for Materials Science (NIMS), 1-2-1 Sengen, Tsukuba, Ibaraki 305-0047, Japan

[5] Department of Statistical Science, School of Multidisciplinary Science, The Graduate University of Advanced Studies (SOKENDAI), 10-3 Midori-cho, Tachikawa, Tokyo 190-8562, Japan

E-mail: (Y.H.) yhayashi@ism.ac.jp, (R.Y.) yoshidar@ism.ac.jp

## Table of Contents

**A. Relationship between thermal conductivity and its standard deviation**

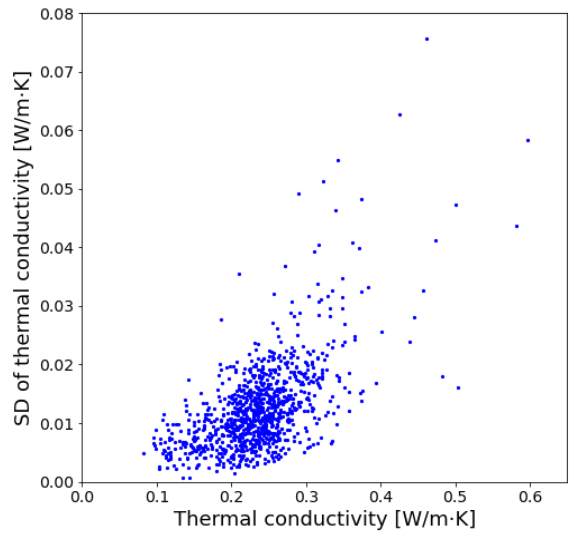

**Figure S1.** Relationship between thermal conductivity and its standard deviation.

## B. Chemical structure of high-density polymers

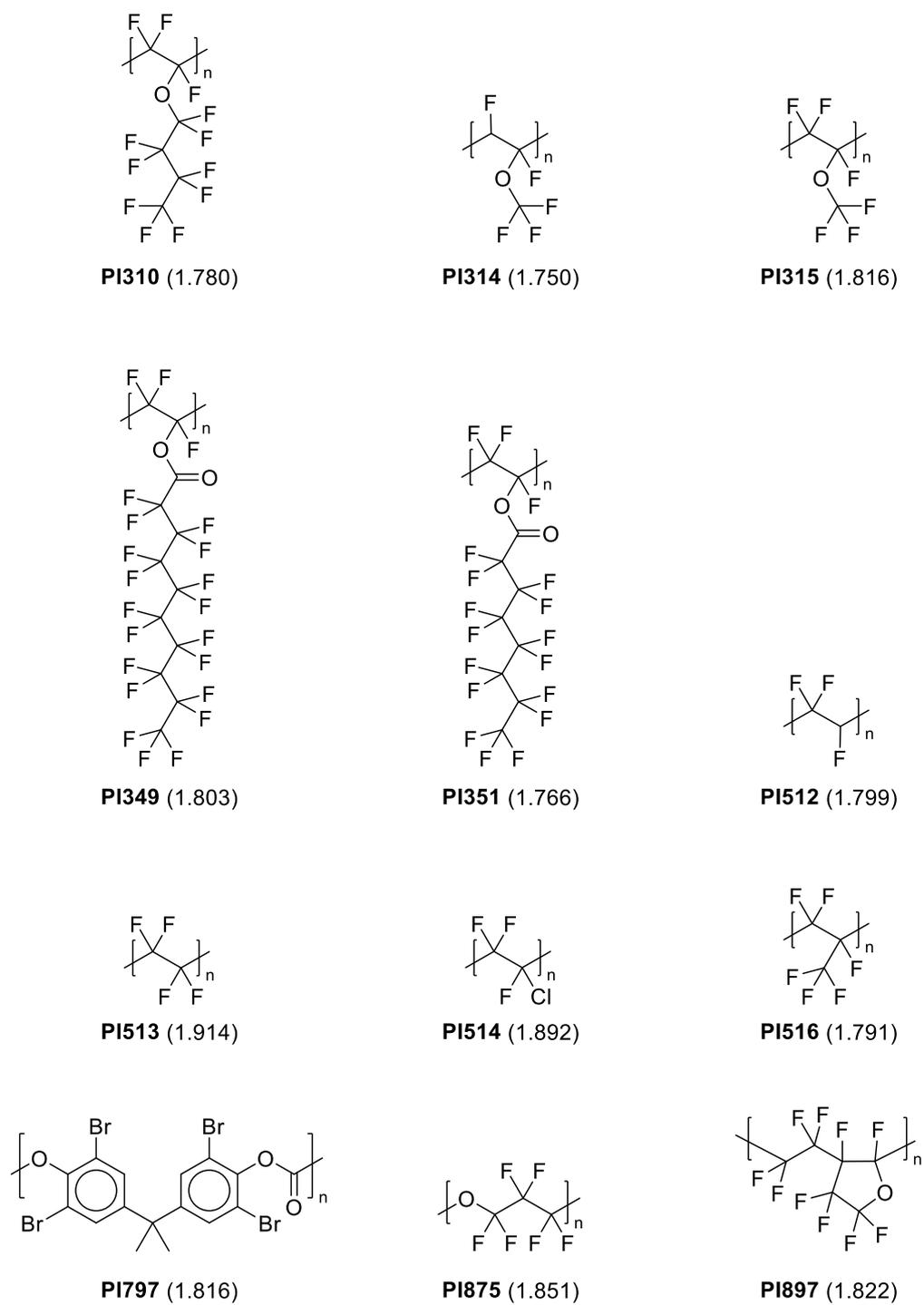

**PI310** (1.780)  **PI314** (1.750)  **PI315** (1.816)

**PI349** (1.803)  **PI351** (1.766)  **PI512** (1.799)

**PI513** (1.914)  **PI514** (1.892)  **PI516** (1.791)

**PI797** (1.816)  **PI875** (1.851)  **PI897** (1.822)

**Figure S2.** Chemical structure of high-density (>1.75 g/cm$^3$) polymers and their calculated density (g/cm$^3$).

## C. Chemical structure of high-refractive-index polymers

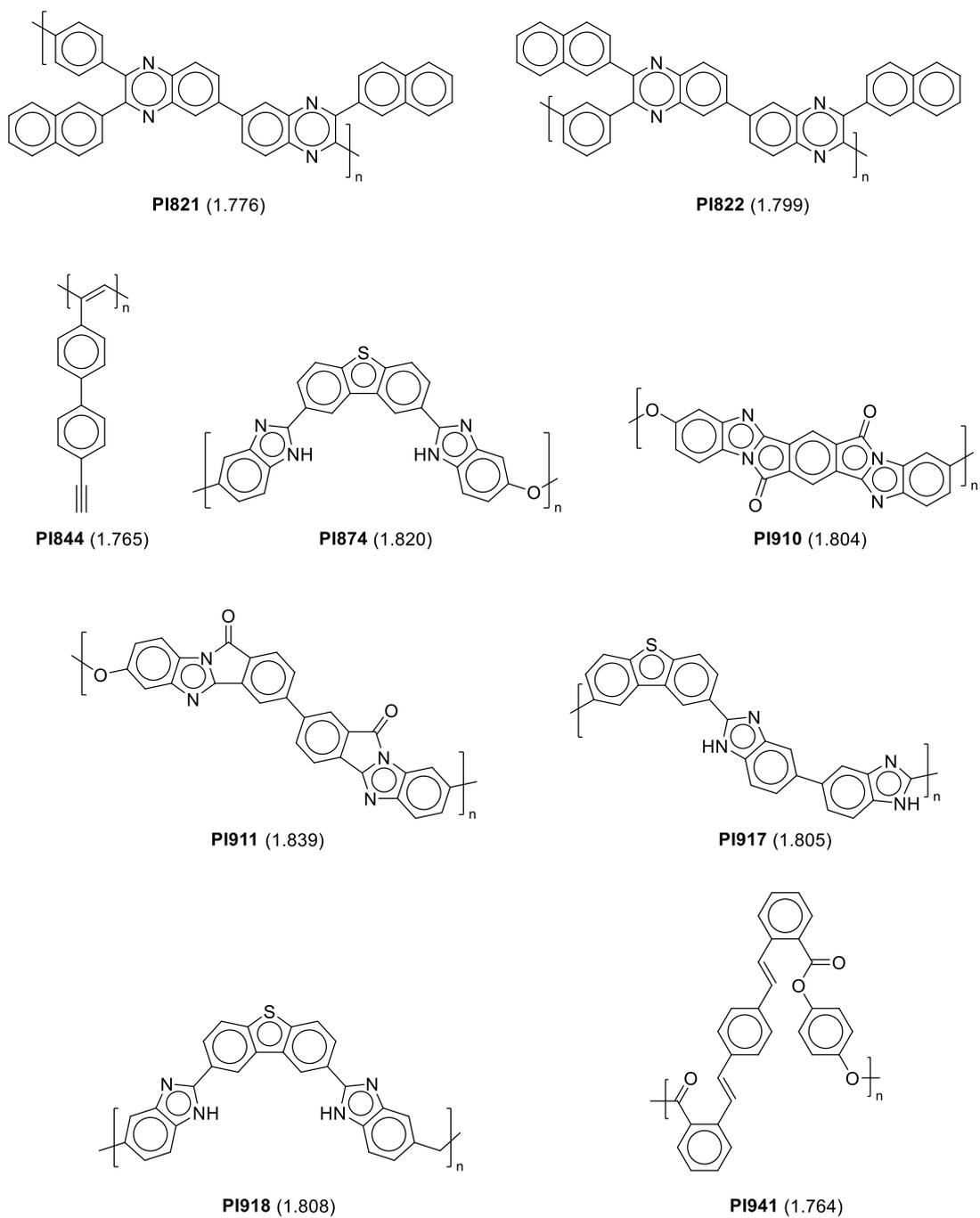

**Figure S3.** Chemical structure of high-refractive-index (>1.75) polymers and their calculated refractive index.

**D. Correlations of mean atomic weights with specific heat capacity and with density**

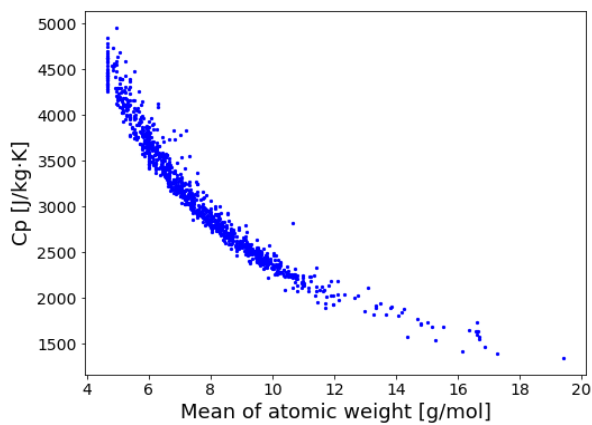

**Figure S4.** Correlation of mean atomic weights with specific heat capacity.

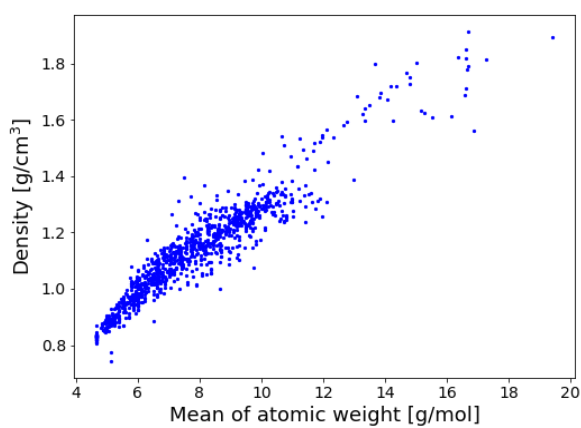

**Figure S5.** Correlation between mean atomic weight and density.